# Curiosity and Pleasure

L. Perlovsky[a], M.-C. Bonniot-Cabanac[b], M. Cabanac[b]

[a]School of Engineering and Applied Sciences, Harvard University; 336 Maxwell Dworkin, 33 Oxford St Cambridge MA 02138; leonid@seas.harvard.edu

[b]Department of Psychiatry & Neurosciences, Faculty of Medicine, Laval University, Quebec, QC, Canada G1K 7P4

Corresponding author: L. Perlovsky, School of Engineering and Applied Sciences, Harvard University; 336 Maxwell Dworkin, 33 Oxford St Cambridge MA 02138; Tel. 781-377-1728; Fax: 781-377-8984; Email: leonid@seas.harvard.edu

Abstract

Heuristic decision making received wide attention due to the work of Tversky and Kahneman (1981) and inspired multiple studies of irrationality of the human mind and a fundamental disregard for knowledge. But what is the source of all human knowledge, including heuristics? We discuss the hypothesis that acquisition of knowledge is a deeply rooted psychological need, a motivational mechanism for perception as well as higher cognition. We report experimental results showing that acquisition of knowledge is emotionally pleasing. The satisfaction of curiosity through acquiring knowledge brings pleasure. This confirms the hypothesis that curiosity or need for knowledge is a fundamental and ancient motivation on a par with other basic needs, such as sex or food. This paper connects curiosity, knowledge, cognition, emotions, including aesthetic emotions of the beautiful, mechanisms of drives, high cognitive functions, minimization of cognitive effort through heuristics, and knowledge maximization. We anticipate our finding to be an important aspect for several classical fields including cognitive dissonance, personality, self, learning, and new directions in cognitive science studying emotions related to acquiring knowledge, personality types in relation to types of knowledge, relating higher cognitive abilities to knowledge-related emotions, and new directions in aesthetics revealing the cognitive nature of the beautiful and music.

# Drives, emotions, and knowledge

Biologists and psychologists have suggested, at least since the 1950s (Harlow, Harlow, & Meyer, 1950; Berlyne, 1960; Festinger, 1957; Harmon-Jones and Mills 1999), that humans and higher animals possess an innate drive for learning. Its primary or secondary role in an organism was not clear. Analyzing mathematical models of learning, Perlovsky (2006) noted that *all* such models use some mathematical mechanism of this drive. Perception, cognition, and an ability to satisfy any instinctual need would not be possible without a primary innate drive for learning. In all mathematical models this drive directs the organism to increase a measure of the correspondence between sensor signals and mental representations of the surrounding world. Correspondingly, he called this measure of the correspondence *knowledge* (Rehder & Hastie, 2001), and suggested that the drive to increase knowledge, the Need for Knowledge, NfK, (Kagan, 1972) is innate and primary. First we discuss NfK, curiosity, and other related mechanisms without differentiating them; later we address fundamental differences.

The idea of NfK may seem controversial and contradicting the body of work initiated by Tversky and Kahneman (1981). This aspect of the paper will be discussed in details later. The NfK, related areas of curiosity, including need for cognition (Cohen, 1957; Cohen, Stotland, & Wolfe, 1955; Cacioppo & Petty, 1982; Cacioppo, Petty, & Kao 1984) are areas of extensive research with thousands of publications. However its motivational status, its relations to primary or secondary drives remain confused and unclear in recent research. "Cacioppo and Petty's… need for cognition… (its) motivational aspect that cannot be… conceptualized as needs, drives, or instincts." (Cacioppo et al 1996).

Loewenstein (1994) emphasized that "Theorizing about curiosity has been largely moribund… the general loss of interest in motivational phenomena such as curiosity… Virtually all… research has examined the cognitive strategies the people use for problem solving… Almost no research on why people are so powerfully driven to solve such problems… even so many researchers… have been struck by the intensity of individual's efforts… in the absence of material rewards… Curiosity involves an indissoluble mixture of cognition and motivation… drive or not a drive is probably unanswerable." In subsequent publications Lowenstein and colleagues concentrated on visceral aspects of curiosity (e.g. Lee, Loewenstein, Ariely, Hong, & Young, 2008). Litman (2005) developed a theory of curiosity related to interest-deprivation and motivational wanting-liking model. Schmidhuber (2009) suggested a theory of curiosity driven by a desire to find a maximally compressed representation of reality. After describing our work, we relate it in more details to the above discussion, and to various existing theories and experimental results.

Here we suggest a motivational theory of curiosity, addressing the fundamental need for knowledge. We briefly summarize discussions that curiosity is fundamentally important for cognition as a whole. We conceptually review mathematical models of the mind-brain, and discuss that increasing knowledge is a part of all artificial intelligence algorithms and mathematical models of cognition. We argue these models describe a psychological drive that is primary and autonomous at lower levels of cognition (such as perception), and that it acquires aspects of conscious curiosity at higher levels of the mind.

We build on a theory of drives and emotions suggested by Grossberg and Levine (1987). According to this theory, satisfaction or dissatisfaction of a drive is perceived emotionally (consciously or subconsciously). At lower levels of everyday perceptions these emotions usually are below the threshold of consciousness; NfK functions autonomously, similar to other bodily needs. For example, stomach functioning could be below a threshold of consciousness as long as it performs its functions as expected, but it immediately rises to a conscious level, if it fails to performs its normal functioning. Similarly, we are not emotionally excited when correctly perceiving an everyday object, such as a chair. However, if normal functioning of perception fails, if NfK is dissatisfied at this usually autonomous functioning level, we could feel such a condition highly emotionally, we are terrified. This is a staple of thrillers. Therefore, specific emotions related to NfK we identify as aesthetic emotions (Perlovsky, 2010a). They are subjectively perceived as harmony or disharmony between mental representations and surrounding world. These emotions are related to curiosity and to aesthetic emotions (Kant, 1790). Kantian arguments have been reformulated in the contemporary language of psychology and knowledge-related emotions of satisfaction or dissatisfaction of NfK have been connected to aesthetic emotions of the beautiful (Perlovsky, 2010a).

The Grossberg-Levine theory of drives and emotions (1987) implies that the NfK drive includes mechanisms of sensor-like organs and neural circuitry, which measures the knowledge, the correspondence between mental representations and reality, and thus indicate to an organism satisfaction or dissatisfaction of the NfK. The candidate brain circuitry has been discussed in (Levine & Perlovsky 2008; Levine 2009). A satisfaction or dissatisfaction of NfK is perceived emotionally. Emotional pleasure is deeply rooted into physiology (Cabanac, 1971; Cabanac, Duclaux, & Spector, 1971; Cabanac & Duclaux, 1970) and has been proposed as the fundamental mechanism of decision-making (Cabanac, 1992; Cabanac & Bonniot-Cabanac, 2007; Perlovsky, 2009). Mathematical models of NfK have predicted that a fundamental mechanism of perception and cognition includes an evolution of mental representations from vague to crisp (Perlovsky, 2001, 2006). This has been confirmed in brain imaging experiments (Bar et al, 2006; Perlovsky, 2009).

The trade-off between maximizing knowledge and an opposite drive to minimize effort, including minimizing cognitive effort through heuristic thinking is discussed in (Levine & Perlovsky, 2008; Levine 2009). Both drives are hypothesized to be evolutionary adaptations. Candidate brain pathways involved in NfK were analyzed in (Levine 2009). Despite the significance of the topic, referenced evidence and theoretical discussions, experimental studies of knowledge-related emotions have been limited, limited evidence exists for a motivational status of curiosity as related to pure knowledge, without visceral effects and without bodily instincts.

Here we address the curiosity for learning new knowledge, and ask, can the pleasure of this effort be demonstrated experimentally? This issue has not been previously studied and it has a principled significance for the NfK hypothesis: is NfK a primary drive caused and rewarded with pleasure? The present study explores the hypothesis that indeed satisfaction of curiosity and acquisition of knowledge is experienced emotionally as pleasure. Remaining difficulties are discussed near the end of the paper.

# Methods

17 women ( 52.9 $\pm$ 17.5 s.d. yr) and 15 men  ( 52.5 $\pm$ 14.7s.d. yr) participated anonymously in the study. After an identification questionnaire: age, gender, level of education, interest in politics, belonging to associations, and marital status, each participant was assigned first to Protocol 1 (P.1) to control for pleasure vs. reading and knowledge independent from curiosity, and second to Protocol 2 (P.2) to explore pleasure *vs*. curiosity. Protocol 1 controlled for pleasure of knowledge, independent from curiosity.  Half the group started with Protocol 1 then Protocol 2, the other half of the group followed the reverse order (P2, then P1). In both protocols, the participant received a list of ten items (statements, questions, or pieces of information), one to explore pleasure vs. learning (P1), and the other to explore pleasure vs. curiosity (P2).  The participants were asked the following:

Protocol 1 (P.1):
- on the answering chart "P.1 knowledge" on ten indented lines the participant would indicate with a YES or a NO, with a pencil mark, whether he/she knew each of the 1-10 pieces of information provided on the knowledge list. The magnitude estimations of previous knowledge about items provided measured this way is abbreviated later as *P1. prev.know*.

**Examples of items from Protocol 1 (Pleasure *vs*. Knowledge) :**
**Item 1: What is the meaning of the word « Huguenot »?**
**Item 3: Do you know how to bake bread?**
**Item 9: Do you know how Van Gogh died?**

the answering chart "P.1 Pleasure of learning" bore ten 140 mm-long horizontal lines; on these lines the participant would rate, with a pencil mark, the intensity of the pleasure/displeasure experienced when reading each piece of information provided on the knowledge list. In the middle of each line a zero indicated absence of any hedonicity (indifference); the right part of the line would indicate positive hedonicity (pleasure) and the left part negative hedonicity (displeasure):

**For each item: on the line below rate the pleasure/displeasure of learning it (after reading the answer)**

-    **---------------------------------0--------------------------------**  **+**

The magnitude estimations of pleasure of reading the answers provided measured this way is abbreviated later as *P1. rtng.pleas.rd*.

Protocol 2 (P.2) consisted of a new list of ten questions, different from those in P.1, about various aspects of life or the environment. The participant would rate on two answering charts his/her curiosity to know the answer, and then the pleasure experienced when receiving the answer to the question:

**Examples of items from Protocol 2 (Pleasure *vs*. Curiosity)**

**Do you wish to know the answers to the following 10 items? Answer y/n**
**Item 1: Do you want to know the meaning of the word  « *Anamorphosis* »?**
**Item 3: Do you want to know who said « The less one    thinks, the more one talks »?**

**Item 9:Do you want to know the criteria used to rank   restaurants?**

**For each item: on the line below rate the pleasure/displeasure of learning it (after reading the answer)**

  - --------------------------------0)------------------------------- +

    - the answering chart "P.2 Curiosity" bore ten 140 mm-long horizontal lines to rate, with a pencil mark, the intensity of experienced curiosity when reading the ten items that would be found on curiosity list.  In the middle of each line a zero indicated no curiosity; the right part of the line would indicate positive curiosity and the left part negative curiosity (rejection); the magnitude estimations of curiosity regarding the items provided measured this way is abbreviated later as *P2. ratg.curios.*
    - the answering chart "P.2 Pleasure of learning" bore ten 140 mm-long horizontal lines to rate, with a pencil mark, the intensity of the pleasure/displeasure experienced when reading each piece of information that answered each of the 1-10 questions raised on curiosity list.  In the middle of each line a zero indicated absence of any hedonicity (indifference); the right part of the line would indicate positive hedonicity (pleasure) and the left part negative hedonicity (displeasure). The magnitude estimations of pleasure of reading the answers provided measured this way is abbreviated later as *P2. pleas.learn.*

## Results

    A first-look at the results is contained in correlation matrixes shown in Tables 1 and 2. For this first-look overview of relationships among all variables, categorial variables (No, Yes) were substituted with (0, 1) correspondingly. In Table 1 correlations were computed within the entire data set (32 x 10 = 320) items for each entry in the table. In Table 2 correlations were computed for each subject separately (10 items) and then averaged over all subjects (32) (within-participant computations); again total of 320 items contributed to computation of each entry in the table. Table 2 could give significantly different results from Table 1 for example, if contributions to correlations come from between subject variations, while within subject data were significantly different (say uncorrelated). This example illustrates that Table 1 gives relevant data for testing our hypothesis about correlation between curiosity and pleasure. In reality both tables show comparable correlation between curiosity and pleasure, lead to the same conclusion and this discussion becomes irrelevant. For completeness, Table 3 shows average and standard deviation values.

    If somewhat arbitrary, we choose significance level corresponding to $p < 0.001$ of accepting the null hypothesis if it is true (no correlation), a significance threshold in both cases is approximately 0.53. It is significant to note that only one correlation in each matrix is significant, that is the correlation between rating on curiosity and on pleasure of learning. This confirms the hypothesis that satisfaction of NfK has a significant hedonic component for subjects with higher rating on curiosity.

| | Gender | Age | P1. prev.know | P1. rtng.pleas.rd | P2. ratg.curios | P2. pleas.learn |
|---|---|---|---|---|---|---|
| Gender | 1 | 0.0146 | 0.0181 | 0.1472 | 0.1638 | 0.1567 |
| Age | 0.0146 | 1 | 0.0532 | -0.0929 | -0.0905 | -0.1443 |
| P1. prev.know | 0.0181 | 0.0532 | 1 | -0.1495 | 0.1527 | 0.1203 |
| P1. rtng.pleas.rd | 0.1472 | -0.0929 | -0.1495 | 1 | -0.0227 | 0.0071 |
| P2. ratg.curios | 0.1638 | -0.0905 | 0.1527 | -0.0227 | 1 | **0.6099** |
| P2. pleas.learn | 0.1567 | -0.1443 | 0.1203 | 0.0071 | **0.6099** | 1 |

Table 1. Correlation matrix computed within the entire data set. For this first-look overview of relationships among all variables, categorial variables (No, Yes) were substituted with (0, 1) correspondingly. A significance threshold here is 0.53, corresponding to p > 0.999 rejection of the null hypothesis (no correlation). Only one correlation is significant, between rating on curiosity and pleasure of learning, confirming the hypothesis.

| | Gender | Age | P1. prev.know | P1. rtng.pleas.rd | P2. ratg.curios | P2. pleas.learn |
|---|---|---|---|---|---|---|
| Gender | 1 | 0.0146 | 0.0181 | 0.1472 | 0.1638 | 0.1567 |
| Age | 0.0146 | 1 | 0.0532 | -0.0929 | -0.0905 | -0.1443 |
| P1. prev.know | 0.0181 | 0.0532 | 1 | -0.3033 | 0.1798 | 0.1219 |
| P1. rtng.pleas.rd | 0.1472 | -0.0929 | -0.3033 | 1 | -0.0913 | -0.1122 |
| P2. ratg.curios | 0.1638 | -0.0905 | 0.1798 | -0.0913 | 1 | **0.5927** |
| P2. pleas.learn | 0.1567 | -0.1443 | 0.1219 | -0.1122 | **0.5927** | 1 |

Table 2. Correlation matrix using within-participant computations.

| | Gender | Age | P1. prev.know | P1. rtng.pleas.rd | P2. ratg.curios | P2. pleas.learn |
|---|---|---|---|---|---|---|
| average | 1.53125 | 52.71875 | 0.5562 | 13.4094 | 13.1937 | 15.0781 |
| st.dev | 0.507007349 | 16.47526861 | 0.5011 | 27.4532 | 35.9099 | 32.0781 |

Table 3. Average and standard deviation values (for gender computation, we arbitrary assign 1 to males and 2 to females).

The fact that ratings on pleasure of reading and pleasure of learning do not significantly correlate can be taken as a further indication of the validity of the results: ratings on curiosity were measured independently from ratings on pleasure from reading and knowledge, when curiosity was not involved (this was the purpose of the first protocol). The correlation matrix in Table 1 gives a sufficient statistical characterization of the data for our purpose. The correlation coefficient of 0.61 between "curiosity" and "pleasure" is equivalent to 37% of variance of each of these variables being explained by the other one.

# Discussion

To understand significance and limitations of the present study, we briefly discuss cognitive theories of emotions additional to those mentioned in to first section.

Curiosity is a complex ability, related to several functions of the mind. Wikipedia (2009) as well as most authors consider it an emotion. Its role in cognition is a subject of long debates. Frijda (1987) developed a theory of emotions in behaviorist tradition, considering emotions as epiphenomena, with the concept of "action tendency" as a focal issue. Emotions are, in this view, tendencies to engage in behavior. He discussed several *basic* emotions. (The Emotions 1986). Grossberg and Levitin (1987) proposed a cognitive theory of drives and emotions; in this theory emotions are neural signals communicating satisfaction or dissatisfaction of drives to decision-making parts of the brain. Johnson-Laird and Oatley (1987) proposed a different cognitive theory of emotions. Emotions are cognitively based states. Complex emotions are derived from a *small number of basic* emotions and arise at junctures of social plans. Ortony, Clore, & Collins (1990) consider emotions developed as a consequence of certain cognitions and interpretations. These authors exclusively concentrate on the cognitive elicitors of emotions, and postulate that three aspects determine these cognitions: events, agents, and objects. Ortony & Turner (1990) questioned the view that there exist basic emotions out of which all other emotions are built, and in terms of which they can be explained; these authors suggested that the notion of basic emotions will not lead to significant progress in the field. They assumed that emotions are reduced to appraisals and other states that are not emotions. Cabanac (2002) considered emotions as common currency among motivational states necessary to make decisions in complex environments. Russell (2003) introduced a notion of core affect as an undifferentiated foundation of emotions. Juslin and Västfjäll (2008) discuss a number of neural mechanisms involved with emotions and different meanings implied for the word 'emotion.' Perlovsky (2001) introduced specific aesthetic emotions related to NfK in the spirit of (Grossberg and Levine 1987), and differentiated them from 'lower' emotions corresponding to bodily instincts. Schmidhuber (2009) suggested a theory of curiosity driven by a desire to find a maximally compressed representation of reality. In several respects this theory is similar to Perlovsky (2001) 'knowledge instinct' or NfK. However, it does not describe recent neuroimaging data about perception mechanisms (Bar et al 2006); the idea of maximal compression, it seems, is narrower, and leads to limited aesthetic ideas vs. (Perlovsky 2010a,b).

Up until this time most authors have discussed only few basic emotions, and the role of huge multiplicity of emotions, especially 'musical' emotions seems mysterious. Perlovsky (2010b) considered a process of differentiation of NfK and emergence of a multiplicity (an almost continuum) of 'musical' emotions. These are differentiated aesthetic emotions, cognitively necessary for reconciliation of cognitive dissonances between any pieces of knowledge, which emerged with evolution of language. In some way musical emotions are differentiation of curiosity.

We have not addressed above in sufficient details differences and similarities among NfK, need for cognition, and curiosity. NfK is very similar to need for cognition. First, NfK is a mental operation in the same way as cognition is. At intermediate hierarchical levels of the mind need for cognition may involve solving puzzles, or thinking through and enjoying intellectual challenges. These efforts lead to and to significant extent consist in improving mental representations (with regard to considered puzzles and challenges). So cognition leads to improved knowledge similar to NfK. (We would emphasize, although it is secondary to the content of this paper that, e.g. solving puzzles, does not consist exclusively in conscious steps but is a combination of conscious and unconscious mental activity, Perlovsky 2001, 2006). Cognition is usually attributed to hierarchical levels above perception. At these higher levels NfK and need for cognition are similar, and their satisfaction or dissatisfaction is experienced emotionally. At lower levels of perception NfK acts autonomously, related emotional neural signals are below the level of conscious registration, and comparison to need for cognition is not applicable.

More problematic is comparison of NfK to curiosity. An excellent illustrative example is some people's curiosity to contents of tabloids. Does this kind of curiosity lead to improved cognitive representations? Are other drives than NfK fundamental to this kind of curiosity? These questions remain open for future studies.

Relations of NfK and heuristics discussed by Tversky and Kahneman (1981) need additional clarifications. This requires analyzing cognitive mechanisms of language and cognition (Perlovsky 2009; Perlovsky & Ilin 2010). Language is learned by about 5 years of age, yet ability to understand and act like adults requires the lifetime. And even at the peak of mental powers few people attain crisp and clear understanding of the entire content of culture stored in language. The given references explain these facts as follows. Children can learn language representations early, without a need for life experience, because language representations exist in surrounding language ready-made. Learning cognitive representations requires life experience (and guidance from language representations). Because of this, not only children but also adults, when talking about areas where they lack direct experience, could rely on language and maintain intelligent conversation without real life understanding. Decision making that relies on cultural knowledge stored in language is called decision by heuristics. Heuristics store wealth of cultural knowledge and could be better than judgments from personal life experience. Nevertheless, heuristics repeat what has already been known, and do not lead to accumulation of new cultural knowledge. Language learning is driven by what Pinker (1994) called 'the language instinct'; it is different from NfK or 'the knowledge instinct' as discussed above. Language instinct involves only language and does not involve life experience. Heuristics advantage of relying on culturally accepted and established 'truths' makes one more certain about his or her decisions. They prevent potentially risky 'original thinking,' but at the expense of refusing to acquire new knowledge. We note that in the 11[th] century Maimonides explained the Original Sin and expulsion from paradise due to Adam's refusal to think originally. By eating from the 'tree of knowledge' Adam acquired the knowledge of heuristics (Levine & Perlovsky 2008).

Bartoshuk, et al. (2005) recently warned that mistakes are made frequently when drawing conclusions from cross modality ratings of intensity as well as of hedonicity (Kubovy, 1999); this issue is avoided in the present study by computing two correlation matrices with essentially the same results. This method gives further strength to the conclusion that pleasure is closely correlated to curiosity, both as a motivation and as a reward.

The present study thus confirms the evidence that pleasure/displeasure takes place as a common currency not only among biological and mental motivations (Cabanac, 1992), but, because mental pleasure has been hypothesized to be different from sensory pleasure (Kubovy, 1999), as well in purely mental conflicts of motivations and in decision making (Ramírez, Bonniot-Cabanac, & Cabanac, 2005). Curiosity, may be added as a correlate with pleasure. Such a result would confirm that satisfying curiosity is rewarding (Eckblad, 1978).

Curiosity seems to be a phylogenetically old motivation that proved selectively advantageous in evolution, animals may have some type of 'need' for sensory change (Hughes, 1997). Epistemic curiosity activates reward circuitry and enhances memory (Kang, 2009). The fact that improving knowledge is a zoologically ancient mechanism (Cabanac, Cabanac, & Parent, 2009), primarily based upon hedonicity and thus universal among humans, might be reflected in the absence of any significant correlation found here (Table I) between hedonicity and age or gender.

With evolutionary emergence of representations in the brain, beginning possibly with Amniotes, mechanisms of perception (Cabanac, 1996) had to adapt mental representations to concrete conditions in the world. It is hypothesized, at that level NfK emerged as a basic mechanism, fundamental for survival, acting autonomously, like digestion. In evolution, with complex hierarchy of brain representations taking place in the human, from perception to abstract concepts, and higher up to ideas of the meaning and purpose of life, NfK has driven evolution of higher cognitive functions (Perlovsky, 2010a).

Knowledge related emotions would potentially influence research in emotional intelligence (Mayer et al, 2001), emotional influence on learning (Levens & Phelps, 2008), cognition and consciousness (Phelps, 2005), self and personality (Luu, Collins, & Tucker, 2000) including personality types in relation to types of knowledge (Bartoshuk, 2010; Mauss & Robinson, 2009; Farb et al, 2010), as well as knowledge-related motivational dimensions of emotions (Harmon-Jones, 2004; Gable & Harmon- Jones, 2010; Perlovsky, 2010a). There are several directions to these future studies including neural mechanisms of the tradeoff between the NfK and cognitive effort minimization or need for closure (Levine & Perlovsky, 2008; Levine 2009), qualities of emotions related to NfK (Perlovsky, 2010b), and personality types with regard to each of these.

We would emphasize that hedonic aspect of curiosity reported in the Results section is only one in a series of arguments establishing curiosity as a primary need. It should be taken together with theoretical arguments and experimental evidence discussed in the first and this last sections of the paper. Results reported here confirm a path discussed in (Levine & Perlovsky, 2008; Levine 2009) to reconciliation between heuristic thinking (Tversky & Kahneman, 1981) and knowledge maximizing thinking (Perlovsky 2001, 2006). Mathematical models of roles of emotions in language, higher cognitive abilities including aesthetic, musical, and sublime emotions (Perlovsky, 2006; 2007; 2009; 2010a; b) suggest that aesthetic emotions refer to knowledge-related experiences. Experimental demonstration of the details of involved mechanisms (Levine 2009), their conceptual and emotional aspects at higher cognitive levels, their differentiation, are directions for future research. By confirming that maximization of pleasure optimizes not only physiology but also mental experience, the reported results made a step in the direction demonstrating the fundamental aspect of the NfK hypothesis: NfK is a primary drive, which satisfaction/dissatisfaction produces pleasure/displeasure. This contributed toward connecting several directions of theoretical explorations.